\documentclass{article}
\usepackage{jheppub}

\usepackage{amsfonts}

\def\tr{{\rm Tr\,}}

\def\bea{\begin{eqnarray}}
\def\eea{\end{eqnarray}}
\def\nn{\nonumber}

\def\lmatrix{\left(\begin{array}}
\def\rmatrix{\end{array}\right)}
\def\gsim{\mathrel{\rlap{\lower4pt\hbox{\hskip1pt$\sim$}}\raise1pt\hbox{$>$}}}
\def\lsim{\mathrel{\rlap{\lower4pt\hbox{\hskip1pt$\sim$}}\raise1pt\hbox{$<$}}}
\def\bi{\begin{itemize}}
\def\ei{\end{itemize}}
\def\msbar{\overline{\rm MS\kern-0.5pt}\kern0.5pt}

\title{ Non-abelian lattice gauge theory \\ with a topological action }

\author[ab]{Daniel Nogradi}
\author[a]{Lorinc Szikszai}
\author[ac]{Zoltan Varga}

\affiliation[a]{Eotvos University, Department of Theoretical Physics, Pazmany Peter setany 1/a, Budapest 1117, Hungary}

\affiliation[b]{Universidad Autonoma de Madrid, Instituto de Fisica Teorica UAM/CSIC and \\ 
Departamento de Fisica Teorica, 28049 Madrid, Spain}

\affiliation[c]{Budapest University of Technology and Economics, Department of Theoretical Physics, \\
Budafoki ut 8, Budapest 1521, Hungary}

\emailAdd{nogradi@bodri.elte.hu}
\emailAdd{szikszail@caesar.elte.hu}
\emailAdd{varga.zoltan@wigner.mta.hu}

\abstract{
$SU(2)$ gauge theory is investigated with a lattice action which is insensitive to small perturbations of the lattice
gauge fields. Bare perturbation theory can not be defined for such actions at all.
We compare non-perturbative continuum results
with that obtained by the usual Wilson plaquette action.
The compared observables span a wide range of interesting phenomena: zero temperature large volume
behavior (topological susceptibility), finite temperature phase transition 
(critical exponents and critical temperature) and also the small volume regime 
(discrete $\beta$-function or step-scaling function).
In the continuum limit perfect agreement is found indicating that universality holds for these topological 
lattice actions as well.}

\begin{document}

\maketitle

\section{Introduction}
\label{introduction}

Universality for lattice gauge theory is generally thought to mean that any lattice action can be used in simulations provided 
(1) it reproduces the desired continuum action in the classical continuum limit, (2) is local, 
(3) has the correct symmetries, (4) the continuum extrapolation is performed. 
Due to asymptotic freedom bare perturbation theory then correctly predicts the approach to the continuum.

The possibility that universality in field theory holds more broadly was first addressed in the non-linear
$O(3)$ model in 2 dimensions \cite{Bietenholz:2010xg}. It was shown that a topological lattice action which is
insensitive to small perturbations of the lattice fields, gives the correct results in the continuum. The topological
nature of the lattice action means that the classical vacuum is infinitely degenerate and hence bare perturbation theory 
can not be set up and is inherently meaningless. The only result that may be obtained from a topological action is
the fully non-perturbative one and it is apparently the same as the one with the usual lattice action of the $O(3)$
model. Hence it seems that requirement (1) above can be dropped and universality still holds. Similar results
were also shown \cite{Bietenholz:2012ud, Bietenholz:2013ria} to hold for the 2d XY model as well. 
\footnote{$U(1)$ gauge theory was studied with a topological action in
\cite{Akerlund:2015zha, Akerlund:2016wpf} but it is trivial in the physically relevant continuum limit.}

From the point of view of
the path integral a useful way of thinking about the type of topological actions we investigate is the following.
Field space is divided into two sets: one, where the action is zero and two, where the action is infinite.
Hence configurations from the former enter with equal weights and fluctuate freely 
while configurations from the latter are forbidden. The
only non-trivial information about the field theory is then encoded in the boundary separating the two sets.

We investigate the same phenomenon for non-abelian gauge theories. A topological lattice action can easily be defined
analogously to the $O(3)$ model and hence bare perturbation theory is again meaningless. Nevertheless we show that for
$SU(2)$ pure gauge theory the non-perturbative continuum results obtained with the topological action agree 
with that of the usual Wilson plaquette action which we know is in the correct universality class of Yang-Mills theory. 
The compared observables are sensitive to a wide range of interesting
physics. At $T=0$ we calculate the topological susceptibility and set the scale $t_0$ by the gradient flow. At $T=T_c$
we compare the dimensionless ratio $T_c \sqrt{8t_0}$ and also the critical exponent $1/\nu$ 
from the Binder cumulant of the Polyakov
loop. In small physical volumes we calculate the discrete $\beta$-function (or step-scaling function) at two values
of the renormalized coupling in the gradient flow scheme. In all cases perfect agreement is found between continuum
extrapolated results using the topological and Wilson plaquette gauge actions. It is worthwhile to point out that a
smooth action that effectively prohibits large plaquettes and is very close to the Wilson plaquette action
for small plaquettes was investigated recently in \cite{Banerjee:2015qtp}.

The organization of the paper is as follows. In section \ref{toplataction} the topological lattice action and the
details of our simulations are given.
Section \ref{topologicalsusceptibility} is dedicated to the topological susceptibility and scale setting, 
in section \ref{deconfinement} the results related to the deconfinement phase transition are discussed 
and in section \ref{smallvolume} the results for
the discrete $\beta$-function are presented. Finally in section \ref{conclusion} we end with a conclusion and possible
future aspects.

\section{Topological lattice action}
\label{toplataction}

We seek an action which is gauge invariant. The simplest possibility is to use the usual plaquette $P$ as the only building
block,
\bea
S &=& \sum_P S(P) \nn \\
S(P) &=& \left\{  
        \begin{array}{ll}
        0      & \mathrm{if} \qquad  1 - \frac{1}{2} \tr P  < \delta \\ 
        \infty & \mathrm{otherwise} 
        \end{array}\right. 
\label{eq:top}
\eea
where the sum is over all plaquettes on the lattice. On a given lattice volume the only parameter is $\delta$ which will
play the role of a bare coupling. Clearly, as $\delta \to 0$ only links close to unity are allowed hence $\delta\to 0$
will correspond to the continuum limit.

Note that the action (\ref{eq:top}) has only two values, $0$ or $\infty$, hence divides the space of links into two
subsets. On one, which contain the unit links, the action is zero and hence the links fluctuate freely without any
weight and contribute equally to the path integral. In particular, the vacuum is infinitely degenerate. 
On the second set of links the action is $\infty$ meaning that links
are forbidden there and contribute nothing to the path integral. The only dynamical information is carried by the
boundary of these two sets defined by $\delta$. As the continuum is approached, $\delta \to 0$, links have less and less
phase space to fluctuate but still always have equal weight. Gauge invariance is encoded in a gauge invariant definition
of the boundary between allowed and not allowed links.

It is well-known \cite{Luscher:1981zq} that if the plaquettes on a lattice are all restricted to be very small,
$1-\frac{1}{2}\tr P < 0.015$, then
a geometric integer definition of the topological charge $Q$ can be given. A slightly more permissive bound was later
derived in \cite{Neuberger:1999pz}, $1-\frac{1}{2}\tr P < \frac{1}{12(2+\sqrt{2})} \simeq 0.0244$. Hence for very
small bare couplings $\delta$ a local algorithm can not change topology. In practice though the values of $\delta$ we
use in this work are much larger, $\delta \sim 0.6 \ldots 0.8$, and we do encounter topology change frequently 
enough in all large volume runs. In
section \ref{smallvolume} the simulations are done in very small physical volumes where of course $Q=0$ but this is not an
algorithmic artifact but rather the consequence of being in the femto world. Even in this case $\delta > 0.3$.

In all runs with the topological action we use a simple Metropolis algorithm whereas with the Wilson plaquette action a
heat bath algorithm. Both Metropolis and heat bath sweeps are accompanied by two to five overrelaxation steps
\cite{Creutz:1987xi}. An allowed
configuration by the topological action may turn into a forbidden configuration by an overrelaxation step, in this case
the step is rejected and the original configuration is kept.

\section{Topological susceptibility and $t_0$ scale}
\label{topologicalsusceptibility}

The first observable we would like to compare in the continuum is the topological susceptibility. The gradient flow
$t_0$ scale \cite{Narayanan:2006rf, Luscher:2009eq, Luscher:2010iy, Luscher:2010we, Lohmayer:2011si, Luscher:2011bx}
is used to make it dimensionless and set the scale, hence we will compare $\chi t_0^2 = \langle Q^2 \rangle t_0^2 / V$.

\begin{table}
\begin{center}
\begin{tabular}{|c|c|c|c||c|c|c|c|}
\hline
$\delta$    &   $L/a$   & $t_0/a^2$ &   $t_0^2 \chi$    &   $\beta$    &   $L/a$   & $t_0/a^2$ &   $t_0^2 \chi$    \\
\hline
\hline
0.8022      & 20   & 1.6739(8)  &    0.00084(3)   & 2.2986 & 20 & 1.572(2)   &  0.00070(5)  \\ 
\hline                                                                                    
0.7411      & 24   & 3.512(3)   &    0.00123(4)   & 2.4265 & 24 & 3.364(3)   &  0.00117(4)  \\
\hline                                                                                    
0.7031      & 32   & 6.106(6)   &    0.00135(6)   & 2.5115 & 32 & 5.843(3)   &  0.00139(3)  \\
\hline                                                                                    
0.6792      & 40   & 8.95(1)    &    0.00148(5)   & 2.5775 & 40 & 8.93(1)    &  0.00152(5)  \\
\hline
\end{tabular}
\end{center}
\caption{The scale $t_0$ and topological susceptibility with the topological (left 4 columns) and Wilson plaquette
action (right 4 columns).}
\label{top_susc_data}
\end{table}

\begin{figure}
\begin{center}
\includegraphics[width=10cm]{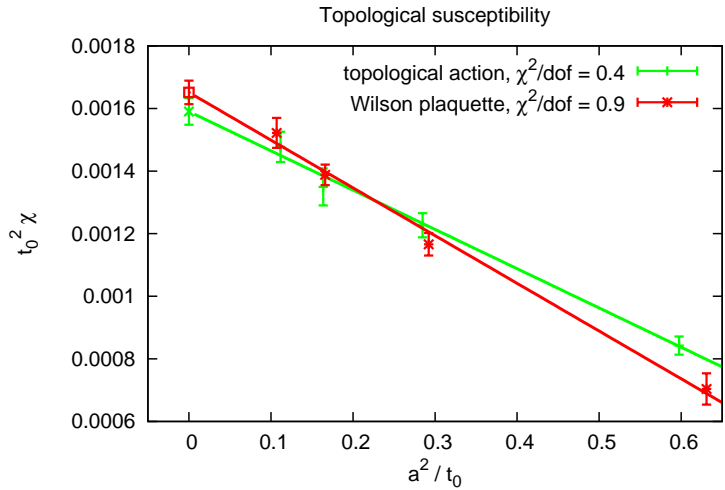}
\end{center}
\caption{Continuum extrapolations of the topological susceptibility.}
\label{top_susc_cont}
\end{figure}

For both the topological action and the Wilson plaquette action the measurement of the gradient flow is done in the same
way, using the plaquette discretization along the flow and the symmetric clover discretization for the observable $E(t)$,
\bea
E(t) &=& -\frac{1}{2}\textup{Tr}F_{\mu\nu}(t)F_{\mu\nu}(t) \\ \nn
\langle t_0^2 E(t_0) \rangle &=& 0.3
\eea
where on the right hand side the choice $0.3$ could in principle be different. Choosing a smaller value would lead to
smaller finite volume effects, smaller errors but larger cut-off effects.
The topological charge $Q$ is also measured along the flow at $t=t_0$ requiring no renormalization for the topological
susceptibility. We confirmed that the continuum results are insensitive to the choice of $t$ as long as it is kept fixed
in physical units. For example, using $t = 0.75 \; t_0 \ldots 1.25 \; t_0$ for the susceptibility leads to identical results.

Finite volume effects are ensured to be negligible within our statistical errors 
by always using symmetric $L^4$ lattices such that $\sqrt{8t_0}/L < 0.25$. Another indicator that finite volume effects
are small is that we always have $T_c L \geq 4$ (see next section).

The measurement of $t_0/a^2$ in our simulations is very precise, its relative error is at least an order of magnitude
smaller than the relative error on the susceptibility. Hence the final errors are completely dominated by the latter.
The data is shown in table \ref{top_susc_data}, the generated number of configurations at each point is $O(10^5)$ with
$O(100)$ configurations separating the measurements.

The continuum extrapolation of $t_0^2\chi$ in $a^2/t_0$
is straightforward and shown in figure \ref{top_susc_cont}. The continuum results
for the two discretizations agree, $0.00159(4)$ and $0.00165(4)$ for the 
topological and Wilson plaquette actions, respectively.

\section{Deconfinement phase transition}
\label{deconfinement}

Next we compare quantities which are sensitive to the deconfinement phase transition which is
second order for $SU(2)$. These can again be compared to the
results obtained with the Wilson plaquette gauge action or with the corresponding quantities in the 3-dimensional Ising
model. First, using the Binder cumulant of the Polyakov loop we will determine the critical exponent $1/\nu$ with
the topological action
and find that it agrees with $1/\nu = 1.5878(4)$ from the 3-dimensional Ising model
\cite{Hasenbusch:1998gh}.
Then the critical couplings are determined on $N_t = 4, 6, 8, 10$ lattices and the dimensionless ratio $T_c \sqrt{8t_0}$
is obtained in the continuum. Again perfect agreement is found between the two actions.

\begin{figure}
\begin{center}
\includegraphics[width=7cm]{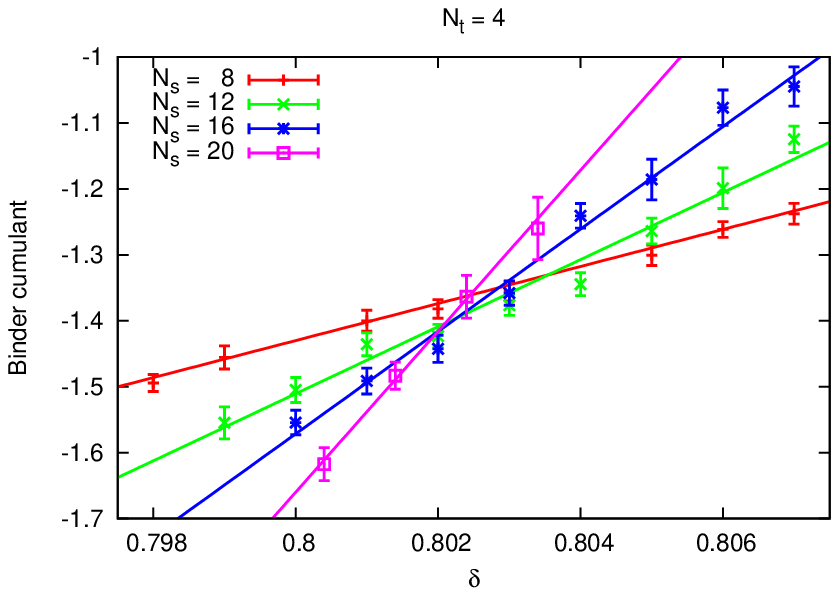} \includegraphics[width=7cm]{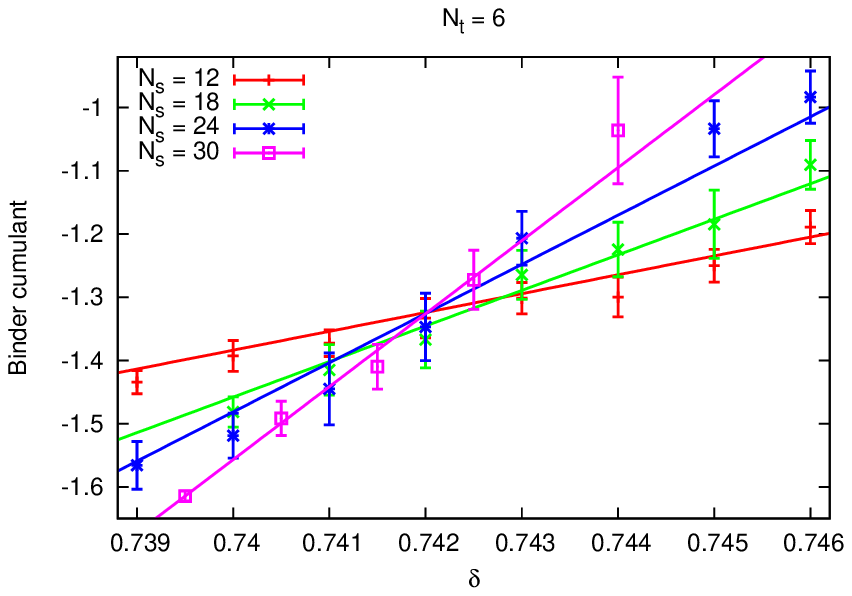} \\
\includegraphics[width=7cm]{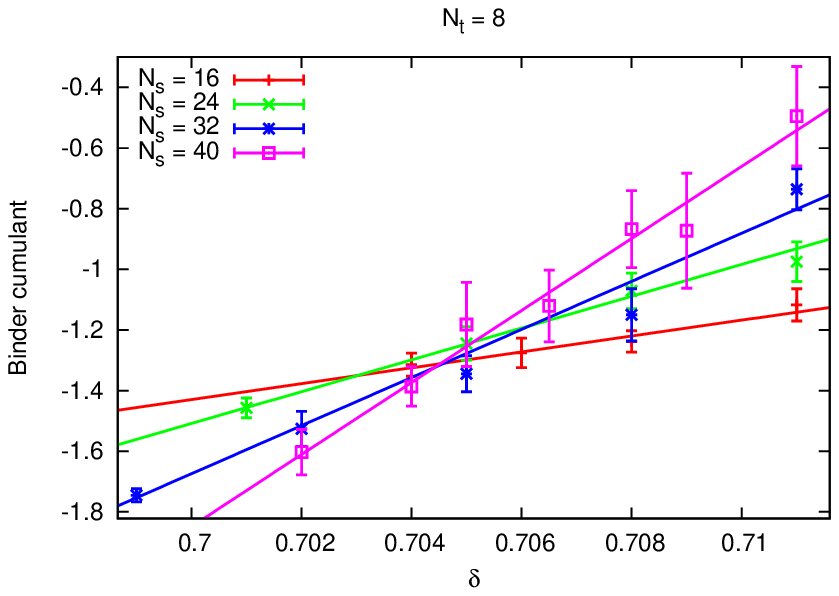} \includegraphics[width=7cm]{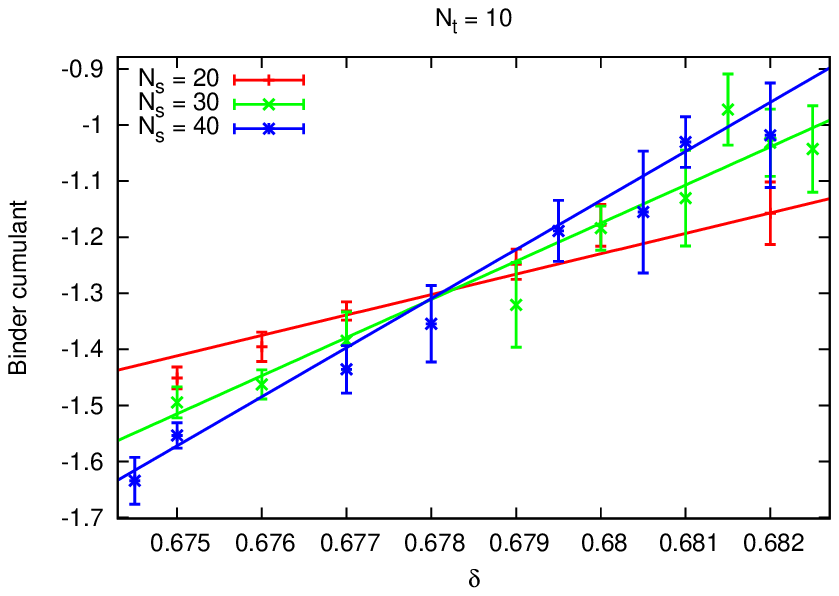}
\caption{The measured Binder cumulants of the Polyakov loop and linear interpolations close to the critical points with the
topological action.}
\label{binderplots}
\end{center}
\end{figure}

We will use standard scaling theory for much of this section; see \cite{Fingberg:1992ju} for more details.
The bare parameters, $\delta$ or $\beta$, are collectively denoted by $\alpha$. 
Since we investigate the deconfinement phase transition at fixed temporal extent, the dependence
on $N_t$ is often suppressed while the spatial volume is denoted by $N_s$. The Binder cumulant of the 
Polyakov loop $L$ is defined by
\bea
g(\alpha,N_s) =\frac{\langle L^4 \rangle}{\langle L^2 \rangle^2} - 3\;.
\eea
The reduced temperature will be denoted by $x = (\alpha - \alpha_c)/\alpha_c$. 
Scaling theory says that close to the critical point
a scaling function $f$ can be defined such that $g(\alpha,N_s) = f\left(xN_s^{1/\nu}\right)$ with some critical
exponent $\nu$. Expanding around $x=0$ we obtain
\bea
g(\alpha,N_s) = f(0) + f^\prime(0) x N_s^{1/\nu} + \ldots
\eea
which means that by considering two spatial volumes, $N_s$ and $bN_s$, for some $b>1$, we may estimate
the exponent as
\bea
\frac{1}{\nu} = \frac{\log\left( \frac{S(bN_s)}{S(N_s)} \right) }{ \log( b ) }\;,
\label{nuexp}
\eea
where $S(N_s)$ is the slope of $g(\alpha,N_s)$ with respect to $\alpha$
close to the critical point. 

Our data for the Binder cumulant for $N_t = 4,6,8,10$ are shown in figure \ref{binderplots} together with linear
interpolations (the largest $\chi^2/dof$ is $1.55$). The generated number of configurations is $O(10^5) - O(10^6)$,
depending on the parameters.
The slopes of these interpolations directly give an estimate of the critical exponent $1/\nu$ via 
equation (\ref{nuexp}) without the need to know the precise location of $\alpha_c$. The results are given in table 
\ref{nutable}.
Clearly, the expected exponent of the 3D Ising model, $1/\nu = 1.5878(4)$, is consistent with our data at each $N_t$ even
at finite $N_s$.

\begin{table}
\begin{center}
\begin{tabular}{|c|c|c|c|}
\hline
$N_t$  &  $N_s$  &   $b$  &  $1/\nu$ \\
\hline
\hline
4      &   12      &  5/3 & 1.71(25) \\
\hline
6      &   18      &  5/3 & 1.40(20) \\
\hline
8      &   24      &  5/3 & 1.60(29) \\
\hline
10     &   20      &  3/2 & 1.54(29) \\
\hline
\end{tabular}
\end{center}
\caption{The critical exponent $1/\nu$ with the topological action from $N_t = 4,6,8,10$ lattices.}
\label{nutable}
\end{table}

Next, we turn to the determination of the critical couplings. Let us denote the intersection of the Binder cumulants
corresponding to $N_s$ and $bN_s$ by $\alpha(N_s, b)$. The dependence on $N_s$ and $b$ is again fixed by scaling theory
and for large enough volumes we have, with some constant $A$,
\bea
\alpha(N_s,b) & = & \alpha_c + A\, \varepsilon( N_s, b ) \nn \\
\varepsilon(N_s, b ) & = & \frac{1}{N_s^{-y_1+1/\nu}} \frac{1-b^{y_1}}{b^{1/\nu}-1} \;,
\label{epsilon}
\eea
where $y_1 = -1$ for the 3D Ising model. Having established that our result for
$1/\nu$ is compatible with the 3D Ising model, we will simply use $y_1 = -1$ and $1/\nu = 1.5878$ in the above
infinite volume extrapolations.

\begin{figure}
\begin{center}
\includegraphics[width=7cm]{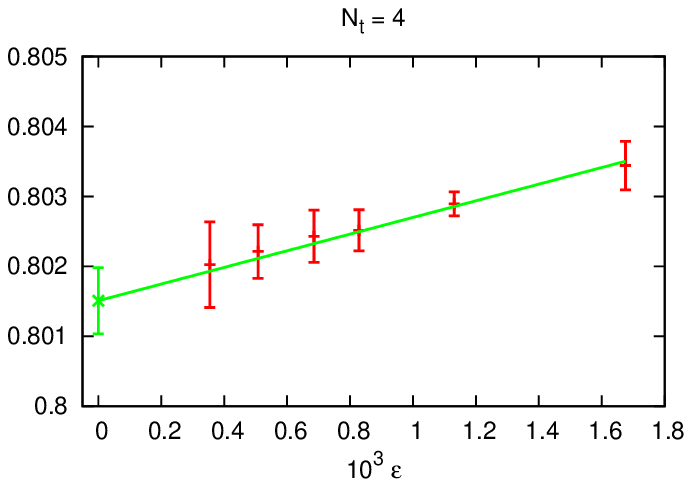} \includegraphics[width=7cm]{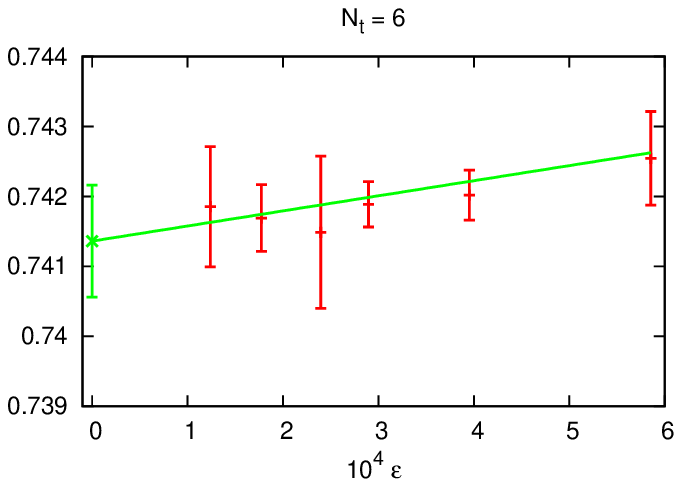} \\
\includegraphics[width=7cm]{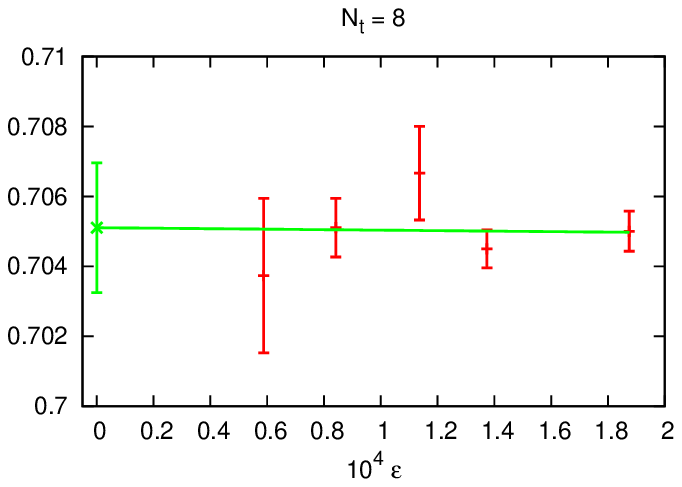} \includegraphics[width=7cm]{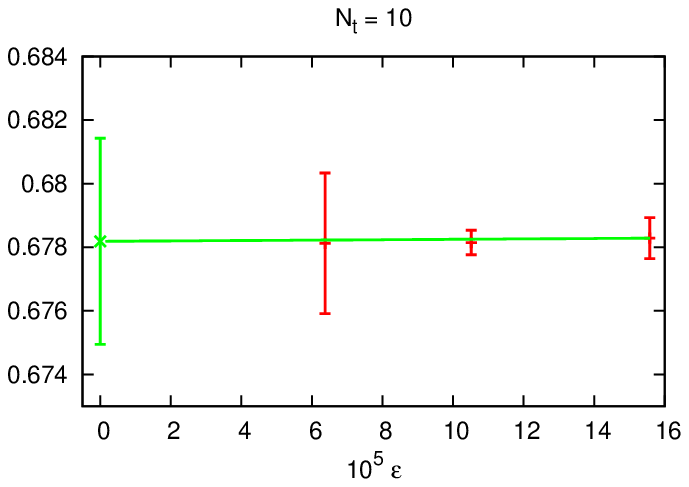}
\caption{Infinite volume extrapolations of the critical couplings with the topological action; see equation
(\ref{epsilon}).
}
\label{epsilonplots}
\end{center}
\end{figure}

The infinite volume extrapolations, using (\ref{epsilon}) is shown in figure \ref{epsilonplots}. Estimating the
statistical error on $\alpha_c$ is non-trivial since the pair-wise intersections for various $N_s$ are correlated;
we use a jackknife procedure starting from the independently measured Binder cumulants.

\begin{figure}
\begin{center}
\includegraphics[width=7.3cm]{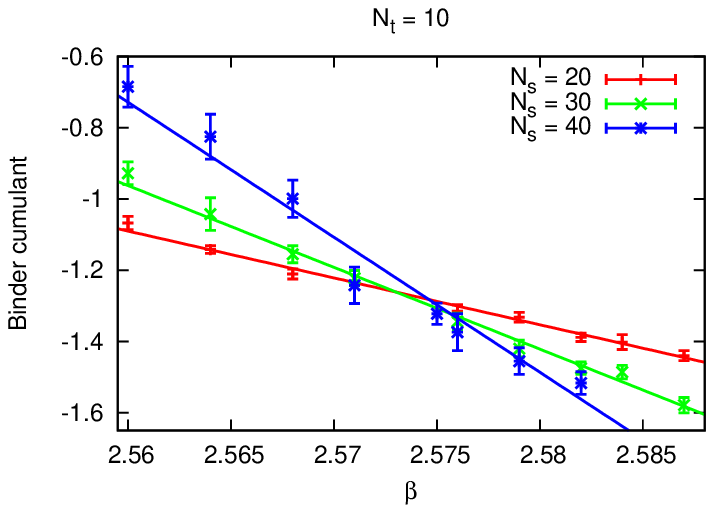} \includegraphics[width=7.5cm]{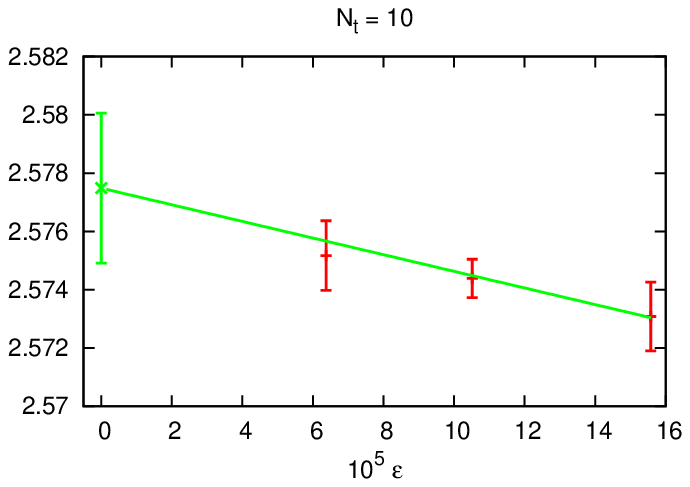}
\caption{Binder cumulants and infinite volume extrapolation of the critical coupling with the Wilson plaquette action
at $N_t = 10$.
}
\label{betaplots}
\end{center}
\end{figure}

The critical couplings with the Wilson plaquette action have been determined in \cite{Fingberg:1992ju} for $N_t =
4,6,8$. Our results for the Binder cumulants and infinite volume extrapolation for $N_t = 10$ is shown in figure \ref{betaplots}.

The final results for the critical couplings are listed in table \ref{crittable} which we subsequently use to
determine $T_c$ in physical units, $T_c \sqrt{8t_0}$.

\begin{table}
\begin{center}
\begin{tabular}{|c||c|c||c|c|}
\hline
$N_t$  &  $\delta_c$ & $t_0/a^2$ &  $\beta_c$ & $t_0/a^2$ \\
\hline
\hline
4      & 0.8015(5)   &  1.68(2) & 2.2986(6)   &  1.572(5)   \\
\hline                                                    
6      & 0.7413(8)   &  3.53(3) & 2.4265(30)  &  3.36(7)    \\
\hline                                                    
8      & 0.705(2)    &  6.05(9) & 2.5115(40)  &  5.8(2)     \\
\hline                                                    
10     & 0.678(3)    &  9.2(3)  & 2.5775(24)  &  8.9(1)     \\
\hline
\end{tabular}
\end{center}
\caption{Infinite volume extrapolated critical couplings and $t_0$ scale for both the topological and Wilson actions. 
The critical $\beta_c$ for $N_t = 4,6,8$ are from \cite{Fingberg:1992ju}. The errors on $t_0/a^2$
contain the error of the critical couplings, that is why these errors are larger than those in
table \ref{top_susc_data}. The 4-volume for the scale measurements are the same as in table \ref{top_susc_data}.}
\label{crittable}
\end{table}

In order to determine the dimensionless combination $T_c \sqrt{8t_0}$ in the continuum
only $t_0/a^2$ needs to be measured at the critical couplings. Even though the
statistical error of $t_0/a^2$ is very small in a given simulation, the uncertainty of the critical coupling itself needs to be
taken into account. By a simple interpolation of $t_0/a^2$ in $\delta$ and $\beta$ we estimate this uncertainty originating
from the uncertainty on the critical couplings. It turns out that this is the dominant source of final uncertainty, 
the statistical error is negligible. In table \ref{crittable} we list the results and the reason for the unusually large error
on $t_0/a^2$, relative to table \ref{top_susc_data}, is the aforementioned effect. Another way of saying it is that if
only the central values of $\beta_c$ and $\delta_c$ are taken, then there is an uncertainty on $N_t$ leading to an
additional uncertainty on $\sqrt{8t_0/a^2} / N_t$ beside $t_0/a^2$.

\begin{figure}
\begin{center}
\includegraphics[width=9cm]{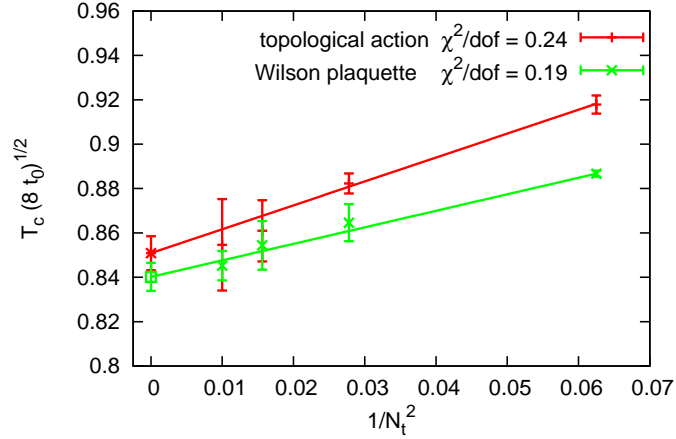}
\end{center}
\caption{Continuum extrapolation of $T_c \sqrt{8t_0}$.}
\label{tccont}
\end{figure}

The continuum extrapolation is shown in figure \ref{tccont}, the results are
$0.851(8)$ and $0.840(6)$ with the topological and Wilson plaquette action, respectively. 
Again, complete agreement is found.

\section{Small volume, perturbative regime}
\label{smallvolume}

In this section we compare quantities in the perturbative regime. The simplest example is given by the discrete
$\beta$-function or step-scaling function \cite{Luscher:1991wu} in small physical volume or femto world
\cite{Luscher:1982ma, Koller:1985mb, Coste:1985mn}
We will use the finite volume gradient flow
scheme with periodic boundary conditions \cite{Fodor:2012qh, Fodor:2012td}. 
In this scheme the renormalized coupling is defined by
\bea
g_R^2(L) & = & \frac{128\pi^2}{9(1+\delta(c))} \langle t^2 E(t) \rangle
\eea
where $c = \sqrt{8t}/L$ is a constant so $g_R^2$ only depends on one scale, $L$, the linear size of the periodic
box. The factor 
\bea 
\label{delta} 
\delta(c) = - \frac{c^4 \pi^2}{3} + \vartheta^4\left(e^{-1/c^2}\right) - 1
\eea 
is such that at tree level the above scheme agrees with $\msbar$, $\vartheta$ is the 3rd Jacobi elliptic function
\cite{Fodor:2012qh,Fodor:2012td}. We set $c=3/10$. The boundary condition is periodic in all directions and it is
well-known that there are degenerate perturbative vacua within this setup. However for small renormalized couplings the
system fluctuates around one of the vacua and tunnelling events are suppressed. In our simulations we work in this
regime and indeed do not detect tunnelling at all as expected.

The expansion of $g_R^2(L)$ in terms of $g_{\msbar}$ is unusual in a periodic 4-torus in the sense that both even and
odd powers can appear, as well as non-analytic logarithms \cite{Coste:1985mn}. 
Particularly for $SU(2)$ purely logarithmically suppressed
terms appear right after tree level; see \cite{Fodor:2012qh} for an extended discussion. 
Hence only the first coefficient of the $\beta$-function is the same in the
gradient flow scheme and $\msbar$. It is important to note that the scheme as such is completely well-defined
non-perturbatively, merely the perturbative expansion behaves in a somewhat unusual way. The $\beta$-function at fixed
$g_R^2$ is a perfectly well-defined and universal quantity. The subscript $R$ will be dropped in what follows.

On the lattice the discrete $\beta$-function, $( g^2(sL)-g^2(L) ) / \log( s^2 )$, or step-scaling function is 
the most easily accessible quantity with a well-defined continuum limit for fixed $g^2(L)$. We set $s=3/2$. It has a
perturbative expansion
\bea
\frac{ g^2(sL)-g^2(L) } { \log( s^2 ) } &=& b_0 \frac{g^4(L)}{16\pi^2} + \ldots 
\label{1loop}
\eea
where, as mentioned above, the terms in $\ldots$ contain both even and odd powers of $g(L)$ as well as logarithms. The
1-loop term $b_0 = 22/3$ is nevertheless universal.

We computed the discrete $\beta$-function at two values of the renormalized coupling, $g^2(L) = 1.5$ and $g^2(L) = 2.5$.
As we will see the first is small enough such that the continuum result is compatible with the 1-loop approximation (\ref{1loop}).
The continuum limit is approached by 3 sets of lattice volumes, $16^4 \to 24^4$, $18^4 \to 27^4$ and $24^4 \to 36^4$ 
corresponding to the scale change $s=3/2$. The desired values of $g^2(L)$ were tuned on the smaller lattices to high
accuracy by tuning the bare couplings of the two actions, $\beta$ and $\delta$, respectively. Then at the same values of
the bare couplings the renormalized couplings were measured on the larger lattices. Finally the results were
extrapolated to the continuum linearly in $a^2/L^2$.

\begin{figure}
\begin{center}
\includegraphics[width=7.5cm]{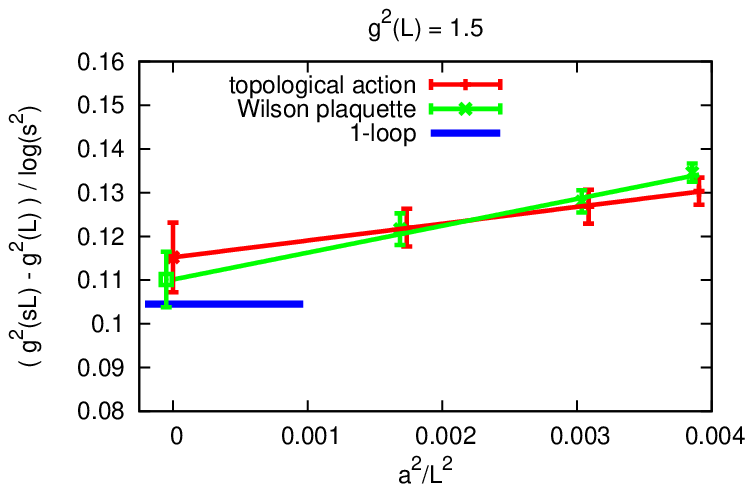}  \includegraphics[width=7.5cm]{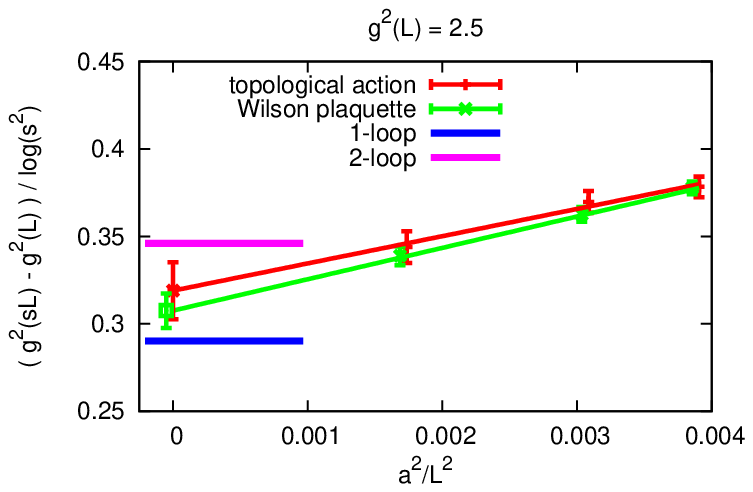}
\end{center}
\caption{Continuum extrapolations of the discrete $\beta$-function at $g^2(L) = 1.5$ (left) and $g^2(L) = 2.5$ (right).
The Wilson plaquette data is displaced slightly for better visibility.}
\label{contbeta}
\end{figure}

The tuned couplings and the measured values on the larger lattices are shown in table \ref{glgsl}.
The continuum extrapolations are
shown in figure \ref{contbeta}, all $\chi^2/dof$ are less than unity. 
As emphasized above the 2-loop result is only shown for orientation, it is not universal
in our scheme. There is perfect agreement for the continuum results between the two actions.
The smaller coupling, $g^2(L) = 1.5$ is expected to be small enough such that
renormalized perturbation theory at 1-loop is trustworthy. This seems to be the case despite the only logarithmically
suppressed terms and the results with the Wilson
plaquette and topological actions are compatible with the 1-loop approximation (\ref{1loop}) 
within $1\,\sigma$ and $1.3\,\sigma$, respectively.
Even though {\em bare} perturbation theory is not possible to set up, {\em renormalized} perturbation theory
behaves as expected.

\begin{table}
\begin{center}
\begin{tabular}{| c || c | c | c || c | c | c |}
\hline
$L/a$ & $\delta$ & $g^2(L)$ & $g^2(sL)$ & $\delta$ & $g^2(L)$ & $g^2(sL)$ \\
\hline
\hline
16 & 0.34174 & 1.498(2) & 1.604(2) &  0.44743 & 2.502(2) & 2.809(5) \\
\hline
18 & 0.33798 & 1.500(2) & 1.602(2) &  0.44049 & 2.499(2) & 2.799(5) \\
\hline
24 & 0.32907 & 1.502(2) & 1.601(3) &  0.42603 & 2.502(3) & 2.781(7) \\
\hline
\hline
$L/a$ & $\beta$ & $g^2(L)$ & $g^2(sL)$ & $\beta$ & $g^2(L)$ & $g^2(sL)$ \\
\hline
\hline
16 & 4.68515 & 1.500(1) & 1.609(1) &  3.65284 & 2.500(2) & 2.806(2) \\
\hline
18 & 4.73710 & 1.500(2) & 1.604(1) &  3.70290 & 2.500(2) & 2.794(3) \\
\hline
24 & 4.86235 & 1.500(2) & 1.598(2) &  3.82180 & 2.502(2) & 2.777(4) \\
\hline
\end{tabular}
\end{center}
\caption{Renormalized couplings, top 3 rows: topological action, bottom 3 rows: Wilson plaquette action.
Left 3 columns: tuned to $g^2(L) = 1.5$, right 3 columns: tuned to $g^2(L) = 2.5$. }
\label{glgsl}
\end{table}

\section{Conclusion and outlook}
\label{conclusion}

In this paper we investigated non-abelian lattice gauge theory with an action that is insensitive to 
small perturbations of the lattice fields. In particular, the classical vacuum is infinitely degenerate.
By comparing continuum extrapolated results to those obtained with the Wilson plaquette action (which is
known to be in the correct universality class) we conclude that even though the topological action
has no classical continuum limit, the quantum continuum limit correctly reproduces the theory.
There are of course only a finite number of comparisons that one can do in simulations but we have chosen
quantities that span a wide range of interesting phenomena. The topological susceptibility, the critical temperature,
critical exponents, and the $\beta$-function in small physical volumes were compared and perfect agreement was found.
This suggests that universality on the lattice is very robust; one is free to modify the lattice action not only with
higher dimensional irrelevant operators but also the relevant operators do not have to be reached in a smooth way
dictated by the classical continuum Lagrangian. 

It may be worthwhile to remember that a positive transfer matrix can not be associated with a topological action at
finite lattice spacing \cite{Creutz:2004ir}. However, as duly pointed out in \cite{Creutz:2004ir}, 
this is not necessarily a problem if the
effect of non-positivity disappears in the continuum limit. In other words if the positivity violation is merely a cut-off
effect. Our results indicate that this is indeed the case since in the continuum the correct Euclidean Yang-Mills theory
is obtained.

A less obvious question concerns the form of scaling violations, $O(a^2)$ specifically, confirmed numerically in our
work. Since bare perturbation theory is not applicable it is not immediately clear that the same scaling is expected as
with the Wilson plaquette action. However, following the same line of reasoning as for the 2-dimensional $O(3)$ model
in \cite{Bietenholz:2010xg}, Symanzik's effective theory, formulated in the continuum, still applies and leads 
to the same $O(a^2)$ scaling violations as with the Wilson plaquette action.

An intriguing application of topological actions might be with dynamical fermions. The Nielsen-Ninomiya no-go theorem
\cite{Nielsen:1980rz}
heavily relies on the classical action, namely that in momentum space the continuum Dirac operator is reproduced
in the classical continuum limit. With a topological fermionic action the classical continuum limit is meaningless.
Hence perhaps the no-go theorem may be circumvented by such actions although it is of course not at all clear how
they would be first defined and then implemented.

\section*{Acknowledgments}

DN would like to thank the Universidad Autonoma de Madrid for hospitality where parts of this work was done.

\end{document}